\documentstyle[amssymb,preprint,aps]{revtex}

\begin{document}

\begin{center}
{\LARGE Electrically Enhanced Free Dendrite Growth in Polar and Non-polar
Systems}{\Huge \smallskip \smallskip \medskip \smallskip }

{\sc Kenneth G. Libbrecht}\smallskip \footnote{%
Address correspondence to $kgl@caltech.edu;$ URL: http://www.its.caltech.edu/%
\symbol{126}atomic/}{\sc , Timothy Crosby, and Molly Swanson }

{\it Norman Bridge Laboratory of Physics, California Institute of Technology
264-33, }

{\it Pasadena, CA 91125\bigskip }

{\small [submitted to the Physical Review Letters, December 19, 2000]}

\end{center}

\noindent {\bf Abstract. We describe the electrically enhanced growth of
needle crystals from the vapor phase, for which there exists a morphological
instability above a threshold applied potential. Our improved theoretical
treatment of this phenomenon shows that the instability is present in both
polar and non-polar systems, and we provide an extension of solvability
theory to include electrical effects. We present extensive experimental data
for ice needle growth above the electrical threshold, where at }$T=-5$ {\bf %
C high-velocity shape-preserving growth is observed. These data indicate
that the needle tip assumes an effective radius} $R^{*}$ {\bf which is
nearly independent of both supersaturation and the applied potential. The
small scale of }$R^{*}${\bf \ and its response to chemical additives suggest
that the needle growth rate is being limited primarily by structural
instabilities, possibly related to surface melting. We also demonstrate
experimentally that non-polar systems exhibit this same electrically induced
morphological instability.\medskip }

\smallskip 

\noindent {\sc Note: if you have trouble reading this document, a robust
.pdf version may be downloaded from http://www.its.caltech.edu/\symbol{126}%
atomic/publist/kglpub.htm\medskip }

The formation of stable spatial patterns is a fundamental problem in the
study of nonlinear nonequilibrium systems \cite{reviews}, and controlling
pattern formation has generated considerable recent interest in light of a
host of possible technological applications. A now-standard example of a
condensed-matter pattern-forming system is the diffusion-limited growth of
free crystalline dendrites, which are nearly ubiquitous products of rapid
solidification, from either liquid or vapor precursors. While the diffusion
equation alone is sufficient to define a relationship between the dendrite
tip velocity and tip radius, typically surface tension (which changes the
surface boundary conditions via the Gibbs-Thomson mechanism) must be
included in order to select a unique needle-like solution. Microscopic
solvability theory has succeeded in furnishing a mathematically consistent
and dynamically stable solution to this problem for simple 2D and 3D
dendrite growth (ignoring, for example, surface kinetic and surface
transport effects) \cite{solvability, saito}. Instabilities and noise
amplification leading to sidebranch generation have also been well studied.

In this Letter we examine shape-preserving needle growth from the vapor
phase in the presence of an applied electrical potential. We previously
described how high electric fields and field gradients near the needle tip
can enhanced its growth, and subsequently drive in a new kind of growth
instability for this system \cite{kglprl, kglpra, comments, mason}. The
growth behavior for small applied potentials is quite well described by an
extension of solvability theory which we describe below. Above a threshold
potential, however, the needle growth cannot be stabilized by the
Gibbs-Thomson mechanism, and so the phenomenon lies outside the realm of
solvability theory. We performed a series of experiments to examine needle
growth in this new electric regime, where we can still observe
shape-preserving growth, but with much increased tip velocities.. From these
measurements we are able to infer the physical mechanisms responsible for
controlling the enhanced needle growth rates.

We consider the case of diffusion-limited growth of a free crystalline
dendrite, which is well known to exhibit a constant-velocity, needle-like
solution with a parabolic tip geometry \cite{saito}. We will assume growth
from the vapor phase, with low solute concentration in a non-polar solvent
gas. When an electrical potential is applied to the growing crystal, the
normal diffusion equation is replaced by the Smoluchowski equation \cite
{kglpra, chandra} 
\begin{equation}
\frac{\partial c}{\partial t}=D\overrightarrow{\nabla }\cdot (%
\overrightarrow{\nabla }c+c\overrightarrow{\nabla }\Phi )
\end{equation}
where $c$ is the solute concentration, $D$ is the diffusion constant, and
the external force felt by the solute molecules is described as the gradient
of the potential 
\begin{equation}
\Phi =-\frac{\alpha }{kT}\left( \overrightarrow{E}\cdot \overrightarrow{E}%
\right)
\end{equation}
where $\alpha $ is the molecular polarizability, and the electric field is
the gradient of the electrical potential $\overrightarrow{E}=-%
\overrightarrow{\nabla }\varphi \ $ \cite{kglpra}. Ignoring interface
kinetics, the continuity equation at the interface yields the normal
component of the surface growth rate $v_{n}=(\widehat{n}\cdot 
\overrightarrow{v}_{surf})$ as 
\begin{equation}
v_{n}=\frac{D}{c_{solid}}\widehat{n}\cdot (\overrightarrow{\nabla }c+c%
\overrightarrow{\nabla }\Phi )|_{surf}
\end{equation}
where $c_{solid}$ is the solid number density and the right-hand side is
evaluated at the solidification front \cite{chandra}. For our experiments
the diffusion length $\ell _{D}\equiv 2D/v_{n}$ is very large compared to
other length scales in the problem, so we can assume the slow-growth limit
where $\partial c/\partial t$ can be taken equal to zero.

The boundary conditions in this problem are defined by the equilibrium vapor
pressure above the crystal surface, which depends on the applied electrical
potential, along with the vapor pressure at infinity, $c_{\infty }$. For a
spherical crystal with radius $R$ a perturbation calculation yields the
vapor pressure 
\begin{equation}
c(R)\approx c_{sat}\left( 1+\frac{d_{0}}{R}-\frac{R_{es}^{2}}{R^{2}}+\frac{%
R_{pol}^{2}}{R^{2}}\right)
\end{equation}
where $c_{sat}$ is the equilibrium vapor pressure above a flat interface, $%
d_{0}=2\beta /c_{solid}kT,$ $\beta $ is the surface tension, $%
R_{es}^{2}=\varepsilon _{0}\varphi _{0}^{2}/2c_{solid}kT,$ $%
R_{pol}^{2}=\alpha \varphi _{0}^{2}/kT,$ and $\varphi _{0}$ is the applied
potential \cite{comments}. The second term in the above expression is the
usual Gibbs-Thomson effect \cite{saito}, the third term arises from the
electrostatic energy of a charged sphere, and the last term is present for
polar molecules, since we are assuming zero field inside the crystal.

With these modifications to the diffusion equation and surface boundary
conditions, we then obtain a modification of the Ivantsov equation relating
the tip velocity, $v,$ and tip radius, $R,$ of a parabolic needle crystal 
\cite{saito, kglpra} 
\begin{equation}
v\approx \frac{2D}{R\log (\eta _{\infty }/R)}\frac{c_{sat}}{c_{solid}}\left[
\Delta _{1}-\frac{d_{0}}{R}+C\frac{R_{elec}^{2}}{R^{2}}\right]
\end{equation}
with 
\begin{equation}
R_{elec}^{2}=\frac{2\varepsilon _{0}}{c_{solid}}\frac{\varphi _{0}^{2}}{%
kT\log ^{2}(\eta _{\infty }/R)}
\end{equation}
where $\Delta _{1}=(c_{\infty }-c_{sat})/c_{sat},$ $C=1+2\gamma \alpha
c_{solid}\Delta _{1}/\varepsilon _{0},$ and $\gamma $ is a dimensionless
geometrical constant. From the spherical growth problem, solved in the limit
of small $R_{elec}$ using a perturbation approach, we expect $\gamma \approx
0.2.$ For $\varphi _{0}=0$ this becomes the slow-growth limit of the
Ivantsov solution if we identify $\eta _{\infty }\approx \ell _{D}.$ This
result supersedes previous treatments \cite{kglprl, kglpra, comments} in
which a cruder approximation was used, which effectively assumed $\gamma =0.$
For ice we find $C\approx 1+4.7\Delta _{1},$ and $R_{elec}\approx 40(\varphi
_{0}/$1000 V) nm \cite{logterm}; thus in our experiments described below the
polarization term is larger than the electrostatic term. For unpolarized
molecules $C=1.$

In the absence of an applied potential solvability theory tells us that the
small capillarity term, $d_{0}/R,$ stabilizes the needle growth; the
decrease in growth velocity for small $R$ brought about by the Gibbs-Thomson
effect opposes the Mullins-Sekerka instability \cite{mullins}. The applied
potential acts to increase the tip velocity for smaller $R,$ thus tending to
destabilize the growth. The electrical perturbation can be incorporated
quite simply into solvability theory by defining a modified capillarity, $%
d_{0}^{\prime }\equiv d_{0}-CR_{elec}^{2}/R,$ which then yields the new
solvability relation 
\begin{equation}
\sigma _{0}=\frac{2d_{0}D}{vR^{2}}\frac{c_{sat}}{c_{solid}}\left( 1-C\frac{%
R_{elec}^{2}}{d_{0}R}\right)
\end{equation}
where $\sigma _{0}$ is the solvability parameter in the absence of an
applied potential. Combining this with the Ivantsov relation then yields a
quadratic equation for the tip radius, $%
R^{2}-R_{0}R+CR_{0}R_{elec}^{2}/d_{0}=0,$ where $R_{0}$ is the tip radius in
the absence of an applied potential. This quadratic equation has no real
roots for $R_{elec}>R_{elec,thresh}=\left( d_{0}R_{0}/4C\right) ^{1/2},$
implying that at above a threshold potential the Gibbs-Thomson effect can no
longer stabilize needle growth.

Ice is a convenient experimental system in which to observe this phenomenon,
and we previously described the behavior of ice needle growth for applied
potentials at or below threshold \cite{kglprl, kglpra, comments}. The
existence of a threshold potential was established, and the tip velocity as
a function of $\varphi _{0}$ below threshold is described by the above
theory (although there remain considerable theoretical uncertainties
associated with the growth of faceted crystals). A principal aim of the
experiments described here is to quantitatively examine the growth behavior
above threshold, beyond the realm of solvability theory.

Our experiments were performed in air at atmospheric pressure, using a
vertical diffusion chamber in which the supersaturation could be changed by
adjusting the temperature profile within the chamber \cite{kglpra}. Crystals
were grown on a thin wire substrate, to which an electrical potential could
be applied. Supersaturation was determined using a combination of diffusion
modeling and frost-point measurements, and we estimate an overall scaling
uncertainty in our reported $\Delta _{1}$ of roughly $\pm $30 percent.

Crystals grown at $T=-15$ C exhibited a classic dendritic morphology with
growth along the $a$-axis, and the tip velocities were observed to increase
approximately linearly with $\Delta _{1}$ as shown in Figure 1(a). Above
threshold at this temperature we nearly always observed the tip-splitting
phenomenon described in \cite{kglpra}. Crystals grown at $T=-5$ C were
needle-like, and at these high supersaturation levels the needle axis was
typically displaced from the $c$-axis, with larger angular displacements
toward the $a$-axis observed for larger $\Delta _{1}$. Fits to these data
using the Ivantsov relation yield tip radii of $R_{0}\approx 1.2$ $\mu $m
and 1.5 $\mu $m for $T=-15$ C and $-5$ C, respectively, independent of $%
\Delta _{1}$. In both cases the data do not follow the trend $v\sim \Delta
_{1}^{2}$ expected from solvability theory, at least over the limited range
in $\Delta _{1}$ shown here. This is probably because the crystals exhibit
some faceting, implying that growth is limited by surface kinetics in
addition to diffusion. Solvability theory does not extend to the faceted
case, but we expect that the additional complication of facets will not
affect our main results to a very large degree.

Above threshold the tip velocities of $T=-5$ C needles increased
substantially, as is shown in Figure 1(c) and Figure 2. The threshold
potential was found to be approximately 1000 volts, independent of $\Delta
_{1},$ implying $R_{elec,thresh}\approx 40$ nm. This is comparable to the
expected $\left( d_{0}R_{0}/4C\right) ^{1/2}\approx 30(1+4.7\Delta
_{1})^{-1/2}$ nm (using $d_{0}=2$ nm and the $R_{0}$ inferred from our
measurements). The quantitative agreement is satisfactory, but again the
observations do not show the supersaturation dependence suggested by our
simple solvability theory. From the supersaturation dependence at $\varphi
_{0}=2000$ volts we find tip radii of $R^{*}\approx 360$ nm using the
Ivantsov relation, again independent of $\Delta _{1}.$ These growth rates
were observed to be the same (to $\pm 15$ percent) for positive and negative
applied potentials.

Adding trace quantities of certain chemical additives to the air in our
diffusion chamber changed the electric needle growth dramatically, as shown
in Figure 1(c). Under certain conditions the electric needles then grew
along the $c$-axis, with velocities approximately four times faster than
without additives, whereas there was no perceptible change to the normal
crystal growth \cite{surfactants}. This chemically and electrically mediated
growth as a function of $\varphi _{0}$ was qualitatively similar to the
behavior shown in Figure 2 scaled by a factor of four, but the
needle-to-needle scatter in the data was considerably greater. These data
imply $R^{*}\approx 90$ nm at $\varphi _{0}=2000$ volts.

With these data in hand we can then examine possible physical mechanisms
responsible for limiting the Mullins-Sekerka instability and producing
shape-preserving growth at a constant tip velocity. Latent heat generated at
the growing tip is a possible stabilizing mechanism, but it appears that the
resulting thermal effects are not very important. Assuming heat generated by
condensation is conducted along the needle and into the solvent gas, then
the temperature rise at the tip is approximately $\Delta T\approx \left(
\kappa _{solid}\kappa _{solvent}\right) ^{-1/2}L\rho vR,$ where $L$ is the
latent heat and $\rho $ is the solid density \cite{kglpra}. Using the
Ivantsov relation for $vR$, this becomes $\Delta T\approx 0.06\Delta _{1}$ ${%
{}^\circ}{}$K$.$ This temperature increase is small, and does not exhibit
the strong $R$ dependence needed to stabilize the tip growth.

Field emission can be examined using the Fowler-Nordheim relation, from
which we estimate a current (for $\varphi _{0}<0$) of $I\approx 14$mA$\cdot
(\varphi _{0}/1000$ V)$^{2}\exp \left[ -310(R/1\text{ }\mu \text{m)(1000 V/}%
\left| \varphi _{0}\right| )\right] .$ Assuming a power $P\sim I\varphi _{0}$
is deposited at the needle tip (likely an overestimate), and assuming the
same simple conduction model as above, we estimate that tip heating becomes
significant for $I\gtrsim 10^{-11}$ A. Ice surface conductance is high
enough \cite{petrenko} that such low currents produce a negligible voltage
drop along a needle.

We measured the current emitted from growing needle crystals, for which
there is a considerable dependence on the sign of the applied potential. For 
$\varphi _{0}=2000$ volts we find $I<$ 1 pA in clean air and $I\approx 1$ pA
for chemically mediated growth (with considerable scatter in the latter
measurements). For $\varphi _{0}=-2000$ volts we measure $I\approx 1-3$ pA
without additives and $I\approx 10$ pA with additives (again with
considerable scatter). When $I\gtrsim 10$ pA we observe that the needle
growth slows substantially, and the tip velocities become quite variable
from needle to needle. We conclude that field emission plays a significant
role when $\varphi _{0}\lesssim -1000$ V and the crystals are grown in the
presence of chemical additives, which results in sizable currents. In all
other circumstances it appears field emission does not limit the needle
growth.

Having eliminated thermal and field emission mechanisms, we propose that the
electric needle growth is stabilized by the diminished structural integrity
of the material when formed into an extremely sharp tip. That is, the needle
tip radius becomes so small that its solid structure is unstable and subject
to deformation driven by surface tension over timescales comparable to the
crystal growth time $\sim R/v.$ There is considerable uncertainty in our
understanding of the structural properties of ice, however, particularly for
such extremely small crystals, which makes detailed calculations difficult 
\cite{petrenko}. Surface melting certainly sets a lower bound on tip radius,
since surface melting is known to produce a quasiliquid layer at these
temperatures, which likely has very little resistance to shear stresses. The
ice quasiliquid layer has a thickness of roughly 20 nm at $T=-5C$ \cite{qll}%
, which is not insignificant compared to the tip radii inferred from our
measurements.

One feature of our data that is readily explained by this model is that the
measured tip radius $R^{*}$ is essentially independent of $\Delta _{1}$ and $%
\varphi _{0}$ in the electric growth regime (at least above a $\Delta _{1}$%
-dependent transition region; see Figure 2). Above the threshold potential
the tip radius decreases and the tip velocity increases to the point that
structural deformations become a dominating influence. This must occur at
some material-determined size scale, so we would expect that the new tip
radius would depend mainly on material properties, and would not depend
strongly on supersaturation or the applied potential, as is observed. The
change in $R^{*}$ in the presence of additives could be explained by the
strongly anisotropic structural properties of ice, since the most rapid
growth observed is for $c$-axis needles. We see that this stabilization
mechanism is a distinct, more direct, manifestation of surface tension.
While normal needle growth is stabilized by the increased vapor pressure
over surfaces with large curvature---the Gibbs-Thomson mechanism---electric
needle growth is stabilized by structural deformations brought about by
surface tension in the sharp tip.

In a separate experiment, we also examined the electric growth phenomenon
for a non-polar molecular solid by examining the growth of iodine crystals
from the vapor. At a supersaturation of $\Delta _{1}\approx 0.1$ we obtained
the results shown in Figure 3. Rapid needle growth above a threshold
potential of $\varphi _{0}\approx 1000$ volts was clearly observed, although
there was considerable scatter in the data. These data provide qualitative
support of the above theory, and clearly verify that this morphological
instability is present in non-polar systems.

In summary, we have examined the detailed physics of electrically enhanced
needle crystal growth from the vapor phase. We find that an extension of
solvability theory explains the existence of a threshold applied potential, $%
\varphi _{thresh},$ and adequately reproduces the magnitude of $\varphi
_{thresh}$ as well as the needle growth behavior for $\varphi _{0}<\varphi
_{thresh}$. For $\varphi _{0}>\varphi _{thresh}$ we observed high-velocity
growth of ice and iodine needle crystals, demonstrating that this new growth
instability is indeed present in both polar and non-polar systems. Extensive
measurements with ice suggest that electric needle growth in this system is
usually not limited by the effects of latent heat deposition or field
emission, but appears to be limited by structural deformation of the needle
tip. Additional experiments with other materials would further elucidate
this process. In particular, since ice is a relatively soft material \cite
{petrenko}, these investigations suggest that harder materials could yield
much sharper tips. Related experiments with refractory metals have yielded
electric needle crystals with nanometer-scale tips \cite{okuyama}, and we
believe that additional research into electrically enhanced growth may lead
to useful applications.


\begin{references}
\bibitem{logterm}  These expressions contain the logarithmic divergence that
is usually present for systems with cylindrical symmetry. The value of $\eta
_{\infty }$ in our experiments was determined by convection currents in our
growth chamber, and was not well constrained. For numerical evaluations we
take $\log (\eta _{\infty }/R)=10.$
\bibitem{surfactants}  We discovered that certain volatile organic
compounds, applied by injecting a small quantity of saturated vapor into the
diffusion chamber, promote $c$-axis growth of electric needles at $T=-5$ C.
We used a ``cocktail'' of VOCs for these experiments, for which we believe
acetic acid was the principal active ingredient. Additive concentrations
were estimated to be roughly at the part-per-million level in air. It
appears that these additives are attracted by electric field gradients to
the needle tip, which greatly enhances their effects compared to growth in
the absence of an applied potential. We are currently studying the details
of this process, and the results will be presented elsewhere.
\end{references}
\end{document}